\documentclass[sigconf]{acmart}

\usepackage{enumitem}
\setlist{  
  listparindent=\parindent,
  parsep=0pt,
  topsep=5pt,
  leftmargin=.45cm
}

\usepackage{subfig}
\usepackage{graphicx}
\usepackage{caption}
\usepackage{svg}
\usepackage{float}
\usepackage{wrapfig}

\usepackage[font=small,labelfont=bf]{caption}

\AtBeginDocument{%
  \providecommand\BibTeX{{%
    \normalfont B\kern-0.5em{\scshape i\kern-0.25em b}\kern-0.8em\TeX}}}

\setcopyright{rightsretained}

\begin{document}
\copyrightyear{2020}
\acmYear{2020}
\acmConference[WebSci '20]{12th ACM Conference on Web Science}{July 6--10, 2020}{Southampton, United Kingdom}
\acmBooktitle{12th ACM Conference on Web Science (WebSci '20), July 6--10, 2020, Southampton, United Kingdom}\acmDOI{10.1145/3394231.3397922}
\acmISBN{978-1-4503-7989-2/20/07}

\title{Misplaced Trust: Measuring the Interference of Machine Learning in Human Decision-Making}

\author{Harini Suresh}
\affiliation{\institution{Massachusetts Institute of Technology}}
\email{hsuresh@mit.edu}

\author{Natalie Lao}
\affiliation{\institution{Massachusetts Institute of Technology}}
\email{natalie@mit.edu}

\author{Ilaria Liccardi}
\affiliation{\institution{Massachusetts Institute of Technology}}
\email{ilaria@mit.edu}

\renewcommand{\shortauthors}{Suresh, Lao and Liccardi}

\begin{abstract}
ML decision-aid systems are increasingly common on the web, but their successful integration relies on people trusting them appropriately: they should use the system to fill in gaps in their ability, but recognize signals that the system might be incorrect. We measured how people's trust in ML recommendations differs by expertise and with more system information through a task-based study of 175 adults. We used two tasks that are difficult for humans: comparing large crowd sizes and identifying similar-looking animals. Our results provide three key insights: (1) People trust incorrect ML recommendations for tasks that they perform correctly the majority of the time, even if they have high prior knowledge about ML or are given information indicating the system is not confident in its prediction; (2) Four different types of system information all increased people's trust in recommendations; and (3) Math and logic skills may be as important as ML for decision-makers working with ML recommendations.
\end{abstract}





\maketitle

\section{Introduction}

Automated decision-aid systems are increasingly prevalent: doctors use diagnostic aids \cite{jiang2017artificial}, judges use risk assessment tools \cite{stevenson2018assessing}, and HR departments use resume screening \cite{ajunwa2016hiring}. Even outside of these specialized domains, everyday web users encounter ML-based recommendations all the time, from search results to targeted ads. In practice, these systems are embedded in larger sociotechnical systems involving human decision-makers and institutional structures \cite{selbst2019fairness,crawford2018anatomy,humanaicollab}. 

Ideally, decision-aid systems should augment human decisions by improving performance when people lack confidence or expertise, but not worsening performance when they would otherwise act correctly. This requires people to be able to decide when to trust the system more than their own judgement, and when to not \cite{lee2004trust}.  Unfortunately, we have seen many cases where ML decision-aid systems have led to worse decisions that can have harmful consequences \cite{skeem2019impact,collins2018punishing,green2019disparate,povyakalo2013discriminate}.  

In order to create ML systems and interfaces that better synergize with humans, it is crucial to develop a robust understanding of how people trust these systems and their recommendations. Providing information about a model and its performance has been proposed as one way to improve appropriate use of ML recommendations \cite{mitchell2019model}, but there have not yet been conclusive empirical evaluations showing if and to whom this sort of information is useful. 

Our work aims to make progress in characterizing how users with different levels of math, logic, and ML knowledge trust ML recommendations. Moreover, we study how this trust is influenced by detailed information about the system and the recommendation. While a growing body of research focuses on issues of trust in ML, and more recently on some aspects of system transparency \cite{yin2019understanding,logg2019algorithm}, to our knowledge this is the first study characterizing trust with respect to both prior knowledge as well as system transparency.




Throughout this paper, we present the design and implementation of a user study to measure (1) people's prior knowledge in both math and logic as well as ML-specific skills, (2) people's trust in both correct and incorrect ML recommendations for a range of questions spanning two different tasks, and (3) how trust changes with information about a model's training data, architecture, performance on a relevant subset, and details about specific predictions.

Our results have important implications for ML model and interface design, ML education, and regulations around system transparency.  In particular, we contribute three key insights: 

\begin{itemize}
    \item People across knowledge levels follow incorrect ML recommendations, even when they would naturally perform the task correctly, and are given information about the recommendation that should suggest it is less trustworthy.
    \item Information about an ML system's training data, model architecture, performance, and recommendation all lead to people following both correct and incorrect recommendations more often. 
    \item Math and logic skills may be as important as basic ML knowledge for human decision-makers working with ML recommendations.
\end{itemize}

\section{Related Work}
The concept of ``trust'' in ML decision-aid systems is measured in different ways, most commonly: ML recommendation agreement \cite{lai2019human,poursabzi2018manipulating,yin2019understanding} and perceived trust in a recommendation \cite{JakeschCHI,kizilcec2016much}. Measuring perceived trust/attitude provides insight into emotions such as fear or bias, but we use recommendation agreement since it provides a more direct measure of actual behavior.

Our work builds on a growing body of research studying similar notions of trust in ML. The review in \cite{lee2004trust} provides a conceptual model of trust in automation and how it varies with context and display characteristics. Some studies have shown a preference for human judgment over algorithmic judgment in ``subjective'' domains such as joke recommendation \cite{yeomans2017making}, after seeing the algorithm make a mistake \cite{dietvorst2015algorithm}, or when the algorithmic output is presented in conjunction with advice from a human \cite{onkal2009relative}.  However, recent work suggests that people may over-trust certain algorithmic recommendations for a range of different advice settings \cite{logg2019algorithm, povyakalo2013discriminate}. \cite{green2019disparate} and \cite{skeem2019impact} specifically point out the harm that can occur when people over-trust algorithmic recommendations that agree with their prior societal biases (e.g., describing how automated risk scoring in the criminal justice system can exacerbate racial and socioeconomic disparities).  \cite{stevenson2018assessing} describes in further detail judges' complicated use of implemented risk assessment tools in the criminal justice system in Kentucky and their impacts over time. 

Other work has focused on how various types of information about a model or recommendation influence trust. Studies have suggested that increased transparency about a recommender system leads to users following its recommendations more often \cite{Sinha:2002:RTR:506443.506619, Cramer2008}. An interesting subset of this work has shown that increased trust can manifest as over-reliance and have negative consequences. For example, \cite{lai2019human} show that more information leads to higher recommendation agreement, even if the information provided is fake. Similarly, \cite{poursabzi2018manipulating} found that participants were less able to correct an inaccurate algorithmic prediction when given the weights of the model and variables used. In the clinical context, \cite{Bussone:2015:RET:2862752.2863267} show how more detailed explanatory information led physicians to trust a clinical decision-aid system more, even when its suggestions were wrong.  The work in \cite{kizilcec2016much} suggests that the relationship between amount of information given and trust is complicated, and that users' trust in a system can begin to decrease if given too much information, particularly when the prediction disagrees with their expectations.
\cite{yin2019understanding} found that higher stated accuracy increases people's trust in a model to some extent, but this may be overriden by seeing its actual performance. However, feedback about a model's actual accuracy is rarely immediate for real-world ML applications (if available at all), so it is useful to focus on the effect of stated performance information as we do in this study, rather than on observed performance or feedback. 

We note that an important and growing body of work focuses on studying interpretability methods for ML models and evaluating their effectiveness with people \cite{tintarev2015explaining,doshi2017towards,DBLP:journals/corr/abs-1902-00006}. Because there is not yet conclusive evidence on which interpretability methods are most useful and understandable for users, in this study we choose to simply state important information about the system and its prediction in standard ways.

In this work, we aim to better characterize task-based trust as well as study the influence of stated information. While other work has focused on narrow categories of given information, we study the effects of four distinct types of ML system information. Moreover, to our knowledge, existing work has not yet studied how trust differs across people with different backgrounds, which we begin to do by measuring how trust varies with math, logic, and ML knowledge. 

\begin{figure*}
\centering
\includegraphics[width=\textwidth]{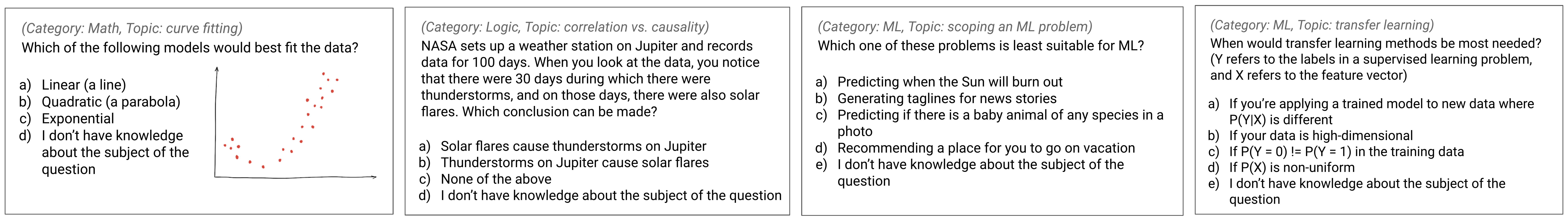}
\caption{Example knowledge assessment questions.} \label{fig:qs}
 \vspace{-1em}
\end{figure*}

\section{Aims \& Challenges}

Our aim in this study is to measure how people incorporate ML recommendations into their decision-making, as a product of (1) the correctness of the recommendation, (2) people's prior knowledge of ML and related math/logic skills, and (3) additional information about the system and recommendation. This study consisted of two main steps: (1) measuring prior math, logic, and ML knowledge through a set of assessment questions, and (2) measuring trust in ML recommendations through a set of task-based questions. Participants sometimes saw these questions along with ML recommendations and/or explanatory information about the system and recommendation (this procedure will be discussed in detail in later sections). There were several key challenges therein:

\begin{enumerate}
    \item \textbf{Accurate assessment of a participant's prior ML, math and logic knowledge}. We asked participants for their perceived ML knowledge and past learning experiences, but self-assessments can be subjective and inaccurate. At the same time, validated instruments for assessing these types of knowledge do not currently exist. 
    
    As a result, we needed to \textbf{design an evaluation of different types of ML, math, and logic knowledge}. This was particularly challenging because the assessment needed to (1) be short, since these questions would be in addition to the main study questions, (2) be distributed across a range of knowledge levels, and (3) encompass not only ML skills but also math and logic skills, since these topics are important ML pre-requisites and would allow us to identify people with strong mathemetical intuition (even if they did not have an ML background).

    
\item \textbf{Effective measurement of human-ML trust.} 
When thinking about how to actually measure people's trust in given ML recommendations, several challenges arise, including:
 
 \textbf{(a) Choosing appropriately difficult tasks}: In order for participants to seriously consider the ML recommendations, it was necessary to select tasks for which people had some intuition but at the same time, would not be easy.
 

\textbf{(b) Presenting real recommendations}: We required all ML recommendations and additional information (architecture, data, performance, prediction, etc.) to be real in order to mitigate the potential confounding factor in participants' behavior if they detected unrealistic data. Therefore, we also needed to pick tasks with open-source data and validated model implementations.

\end{enumerate}

\section{Data and Material Creation}
Our study required the creation of three distinct materials: (1) Assessment questions to measure knowledge of math, logic, and ML, (2) A set of task-based questions to measure trust with real ML recommendations, and (3) Explanatory material about the ML systems and recommendations for each task/question. 


\subsection{Knowledge Assessment}
In step (1) of our study, we assessed participants' knowledge in math, logic and ML. To do this, we created 22 knowledge assessment questions split into 3 categories: Logic (3 questions), Math (8 questions), and ML (11 questions). The questions sampled important topics from each area, ranging in difficulty from concepts that are common knowledge or secondary education-level knowledge to those that require advanced domain knowledge. We chose Math and Logic topics that were particularly relevant to ML. 

We created our own assessment because a validated instrument for assessing ML knowledge does not currently exist. We iterated all the questions starting from a number of concepts identified in 2 instances of an ML course developed by one of the co-authors \cite{LAO2019ADE}. The Logic questions covered the following concept areas (parentheticals indicate the question ID, which will be used to refer to particular questions later on): \textit{correlation and causation} (L1), \textit{decision tree logic} (L2), and \textit{set relationships} (L3). 

The Math questions covered the following concept areas: \textit{curve fitting} (M1), \textit{variance of a sample} (M2),  \textit{probability} (M3), \textit{linear functions} (M4), \textit{means and medians} (M5), \textit{graphical models} (M6), \textit{false positives and negatives} (M7), and \textit{points on a hyperplane} (M8). 

In order to capture knowledge of individuals familiar with different branches of ML, the ML questions covered more concept areas than the previous categories: \textit{describing ML} (ML1), \textit{scoping an ML problem} (ML2), \textit{supervised and unsupervised learning} (ML3), \textit{training vs. testing data} (ML4), \textit{SVMs/kernel functions} (ML5), \textit{gradient descent} (ML6), \textit{k-means clustering} (ML7), \textit{transfer learning} (ML8), \textit{regularization} (ML9), \textit{optimization} (ML10), and \textit{topic modeling} (ML11).

The questions were multiple choice, and all included the option of ``I don't have knowledge about the subject of the question''. Some example questions are shown in Fig. \ref{fig:qs}.

\subsection{Task-Based Dataset} 
In step (2) of the study, participants answered questions with or without ML recommendations. We selected two different applications that currently have real ML implementations: (1) telling apart images of similar-looking animals, and (2) comparing the numbers of people in photographs of crowds. Both applications were centered on image recognition tasks to limit variations based on data modality, and because image-based tasks tend to be faster and simpler for people as opposed to tasks that require interpreting text or numerical data. We used a within-subject study design to test four different treatments (described in Section 5). A participant would encounter each treatment four times (2 with correct and 2 with incorrect recommendations). We selected four comparable sub-tasks within each application to avoid confounding and learning effects resulting from having previously seen an image with another treatment.

For the animal identification application, the subtasks involved identifying an image of an animal out of two similar-looking types of animals: (1) leopards and jaguars, (2) wild boars and hogs, (3) beavers and marmots, and (4) bullfrogs and tailed frogs.  For the crowd comparison application, the subtasks involved telling apart which image out of two images had more people, from sets comprising of: (1) pictures around Venice plazas, (2) pictures from the streets of Shanghai, and (3) and (4) pictures of large crowds scraped from the web, which differed in crowd density.  

\subsubsection{ML Models}
Participants sometimes received ML recommendations and system information for the task-based questions. All ML-related information was real to avoid introducing a confounding factor into participant behavior. We describe the models and datasets that we used below.

\textit{Animal Identification Application:} We used a ResNet50 neural network \cite{he2016deep} pre-trained on the ImageNet dataset \cite{imagenet_cvpr09} implemented in Keras \cite{chollet2015keras} for all subtasks. Although the direct output of ResNet50 is a probability distribution over 1000 classes, we generated performance metrics specific to our binary classification subtasks. We calculated the binary accuracy for each subtask using the 50 ImageNet validation images per class. For each of the 100 validation images per subtask (which spans 2 classes), if ResNet50 predicted that the probability for the correct class was higher than that of the incorrect class, it was counted as a correct classification. The model's performance on the subtasks ranged from 72\% to 89\%.

\textit{Crowd Comparison Application:} We used an implementation \cite{unofficialmcnn} of the multi-column convolutional neural network (MCNN) architecture trained using part A of the ShanghaiTechDataset \cite{zhang2016single}
. The subtasks consisted of 4 sets of crowd images that looked distinct and were sourced from various datasets: (1) UCF-QNRF, consisting of images of extremely large crowds scraped from the internet \cite{idrees2018composition}, (2) Venice, consisting of images from cameras around a plaza in Venice, Italy \cite{Liu_2019_IROS, Liu_2019_CVPR}, (3) Shanghai A-test, a subset of the ShanghaiTechDataset consisting of images of crowds scraped from the internet (which differs significantly from UCF-QNRF in crowd density), and (4) Shanghai B, a subset of the ShanghaiTechDataset consisting of images from cameras around Shanghai streets. Although MCNN outputs a predicted count of people in an image, our subtasks are binary classification problems asking which of two images has more people. To calculate subtask accuracy, we used the model to predict the crowd counts for 100 random, non-repeating pairs of images from the subtask dataset. On these 100 images, if the model predicted a higher number of people for the image that actually had more people, this was counted as correct. The model's performance on the subtasks ranged from 72\% to 93\%.

\subsubsection{Question Selection} 
We selected each question by hand in order to check that the difficulty of the questions were comparable and that the images were reasonable (e.g., not watermarked, not offensive). Moreover, since some treatments showed explanatory information such as the ML model's outputted probablities or counts, these values also needed to be comparable for the chosen examples. 

We observed that examples with incorrect recommendations fell into two categories: (1) those that appeared normal both in the image and the recommendation (e.g., high probability for a single answer, image not dark or blurry), and (2) those that appeared abnormal either because of the image (e.g., dark, blurry) or because of the model's output (e.g., extremely low probabilities/close values for both classes). We expected that trust would vary across these different types of errors, so we chose two incorrect recommendation questions (one ``normal'' and one ``abnormal'') for each subtask, and two correct recommendation questions (both ``normal'').

For the animal identification application, the ``abnormal'' incorrect recommendation questions were those where the ML system outputted very low and very close probabilites for both classes. For the crowd comparison application, the ``abnormal'' incorrect questions were those where one or both of the crowd images were dark, blurry, or contained artifacts.  

\begin{figure}
 \centering
 \includegraphics[width=0.8
 \columnwidth]{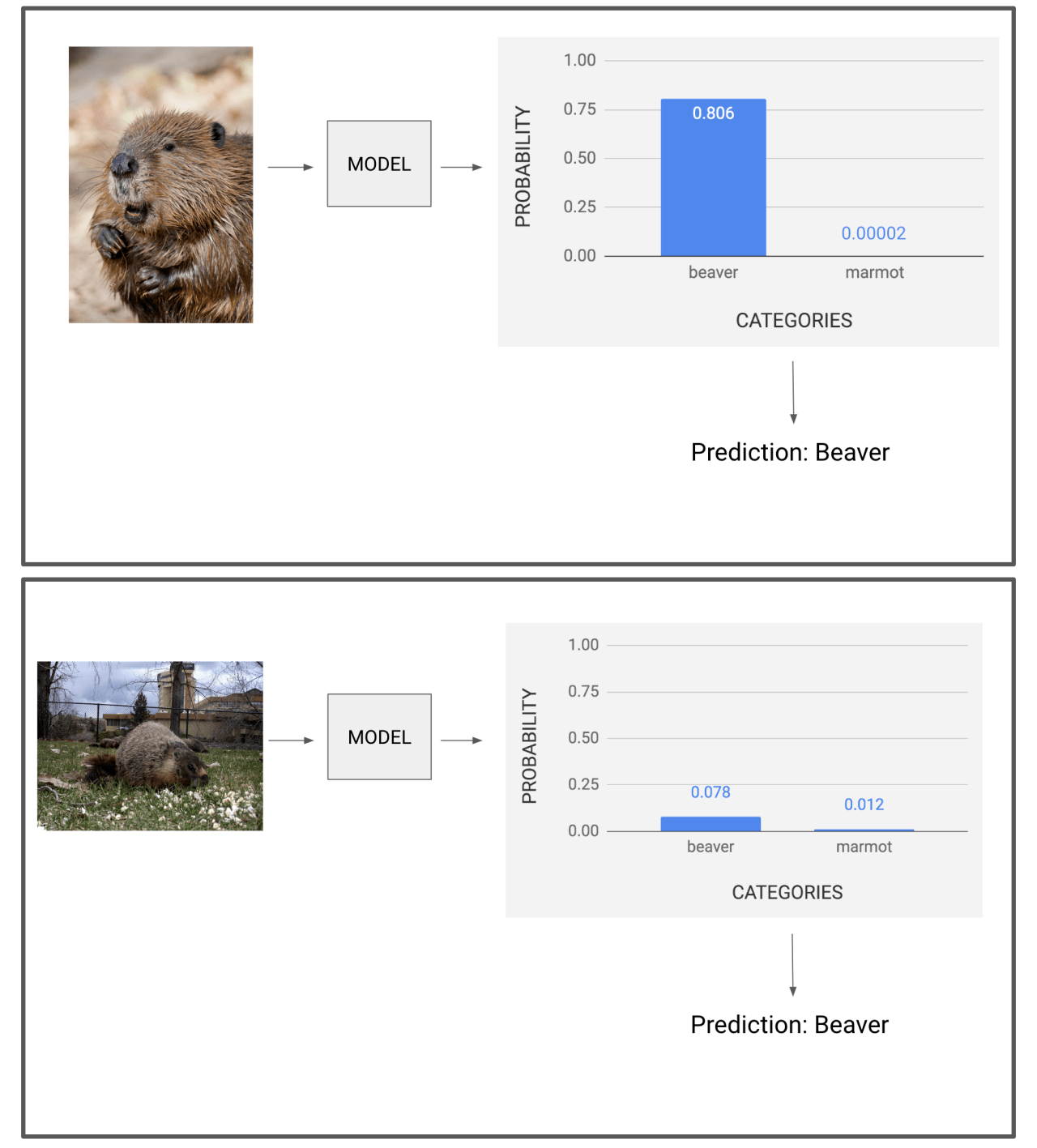}
  \caption{Example screenshots from prediction videos, both normal (top) and abnormal (bottom).}~\label{fig:vids}
\vspace{-1.5em}
\end{figure}

 

\subsection{Explanatory Video Material}

In step (2) of the study, participants sometimes received explanatory information about the ML system and recommendation. Videos with voiceover were used to convey the information as opposed to text or static graphs/tables, since the usage of videos as an instructional tool has been shown to significantly improve recall of conceptual information and creative problem solving over other mediums, particular for lower prior-knowledge participants \cite{mayer1990}. Additionally, the video medium allowed us to track the amount of information the participant received through view times.

    

The videos covered four types of information: (1) model architecture, (2) data used to train the model, (3) the model's performance on a relevant subset of the data, and (4) prediction information for a given question (the predicted probabilities of each class for animal identification, or the predicted counts of each image for crowd comparison).

We created 2 data videos (1 per application), 2 model videos (1 per application), 8 performance videos (1 per subtask), and 32 prediction videos (1 per question)\footnote{All videos are available to view at https://github.com/harinisuresh/ml-trust-materials}. Videos ranged from 15 seconds to 45 seconds. We limited the types of information provided to four to keep the total length of videos for any given question under 2 minutes. Standards for model reporting are still in early stages of research \cite{mitchell2019model}, so these four concepts were chosen based on current norms for describing ML work. The videos were also constructed such that even people without prior knowledge would be able to understand at least part of the information. For example, when the model video describes the architecture for the animal identification model, we included high-level intuition and background about the model (e.g., describing a convolutional neural network as ``passing parts of the image through various filters to learn patterns''). Each video was displayed with English subtitles.

\section{User Study}
This study was accepted by our institution IRB board, and participants were asked for consent prior to the study. 

\subsection{Treatments}
Participants saw the knowledge assessment questions followed by the task-based questions. Each task-based question was associated with one of four treatments:
\begin{enumerate}[topsep=0pt,itemsep=-1ex,partopsep=1ex,parsep=1ex]
    \item \textbf{NoRec}: There is no recommendation provided.
    \item \textbf{RecOnly}: The ML recommendation is displayed with no additional information about the system.
    \item \textbf{RecOptVid}: The ML recommendation is displayed and four categories of information about the system and results are available as videos. Participants are free to view any or none of those videos.
    \item \textbf{RecForceVid}: The recommendation and additional information are available as in RecOptVid. Participants are required to view all videos at least once.
\end{enumerate}

Every participant saw all eight subtasks, and each of the four subtasks within an application had a different treatment that was applied for all questions in the subtask. The assignment of treatment to subtask was random. Because some of the information was shared between subtasks, we showed all the subtasks from a particular application in succession.  The order of the applications, the order of the subtasks within the application, and the order of questions within the subtask were random. 

\subsection{Procedure}
Each participant was given an account on a web application for the study. After logging in, participants saw the following:
\begin{enumerate}
    \setlength{\itemsep}{0pt}
    \item An introductory text describing the types of questions, expected study length, and audio component for the videos.
    \item The knowledge assessment questions.
    \item A 2-minute training video walking through the task-based questions and the different treatments.
    \item Three comprehension-check questions about the training video that must be answered correctly to proceed.
    \item The task-based questions. For each application, and for every subtask in the application (randomly ordered):
        \begin{itemize}
            \item A training page with three examples of problems in the subtask with their ground-truth answers.
            \item Four subtask questions (randomly ordered); recommendations and explanatory videos shown depending on the treatment for the subtask.
        \end{itemize}
\end{enumerate}

To check for validity, each participant was given 8 repeated questions: 3 in the knowledge assessment section and 5 in the task-based section (randomized but limited to one per subtask).  Participants needed to answer a majority of these repeated questions with the same answer as the original question to be considered valid. Each participant saw 22 knowledge assessment questions, 32 task-based questions, and 8 repeated questions for a total of 54 questions.

\begin{figure}
    \centering
    \includegraphics[width=\columnwidth]{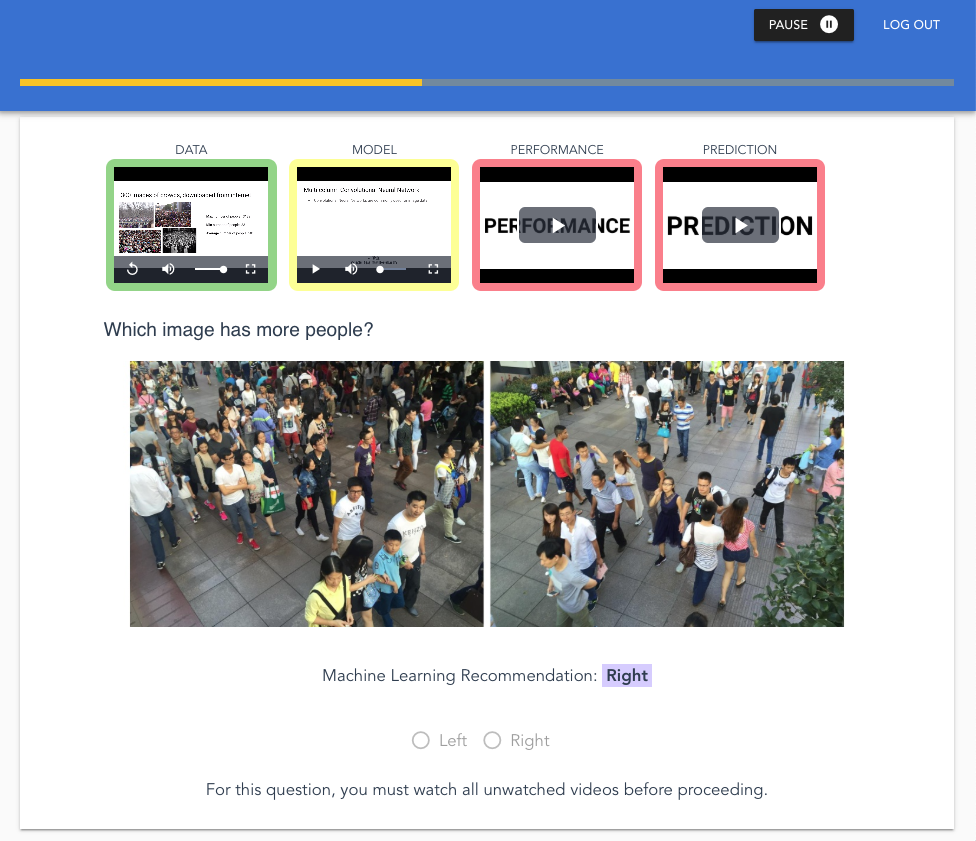}
    \caption{Screenshot of a question as delivered by the webapp for the study. This question has the RecForceVid treatment. In treatment RecOptVid, the videos appear in the same way, but participants are able to click an answer and proceed without watching any videos. In treatment RecOnly, only the recommendation appears, and in treatment NoRec treatment, only the question and choices appear. }~\label{fig:RecForceVid}
     \vspace{-1.5em}
\end{figure}

\subsection{Apparatus}

We built a web application to deliver the study. Each question appeared on a separate page with the image (if applicable) and radio buttons for answer choices. The study immediately proceeded after an answer click was detected to mitigate click fatigue. Participants could pause the study. 

If a question had the RecOptVid or RecForceVid treatments, all 4 videos appeared above the question. Unwatched videos had a red border, videos that had been partially watched in a previous or the current question had a yellow border, and videos that had been watched in their entirety in a previous or the current question appeared with a green border. A sample screenshot of the study is shown in Fig. \ref{fig:RecForceVid}. 

Videos immediately expanded to fullscreen when clicked and automatically paused when minimized. This prevented users from viewing anything else on the screen while the video was playing, or from playing multiple videos at once. Users could not seek forwards past their furthest watched time (but seeking backwards was allowed). For questions in treatment RecForceVid, the answer options were unclickable with the text ``For this question, you must watch all unwatched videos before proceeding'' until all videos were watched fully.

\section{Participants}
We recruited participants for this study by posting on public email lists, Reddit, and Facebook groups.  Interested parties filled out a Qualtrics survey that collected demographic information and a contact email.  We also asked people to self-rank their ML knowledge on a 5-point scale.  From the initial survey, we selected participants based on demographic characteristics and self-reported expertise in order to maintain a diverse population. We created accounts for each selected participant which were delivered via email with a link to the study website.

In total, 175 participants successfully completed the survey and passed the repeated question check.  Gender, ethnicity, age and education level breakdowns are presented in Tab. \ref{tab:demo}.

\begin{table}
\centering
\small
 \begin{tabular}{l l c} 
 \hline
 \textbf{Gender} & \textbf{Num.} & \textbf{Percentage} \\ 
 \hline
 Male & 86 & 49.1\% \\ 
 Female & 86 & 49.1\%  \\
 Other & 2 & 1.1\%  \\
 Prefer Not to Answer & 1 & $<$1\%  \\
 \hline
 \textbf{Ethnicity} & &  \\ 
 \hline
 White & 81 & 46.3\% \\
 Aisan/Pacific Islander & 37 & 21.1\% \\ 
 Black or African American & 25 & 14.3\%  \\
 Hispanic or Latinx & 15 & 8.6\%  \\
 Native American or American Indian & 9 & 5.1\%  \\
 Other & 4 & 2.3\% \\
 Prefer not to answer & 4 & 2.3\% \\
  \hline

  \textbf{Age} & &  \\ 
 \hline
 18-24 & 73 & 41.7\% \\
 25-34 & 39 & 22.3\% \\ 
 35-44 & 51 & 29.1\%  \\
 45-64 & 12 & 6.9\%  \\
 \hline
 \textbf{Highest Education Level} & &  \\
 \hline
 Less than high school diploma & 1 & $<$1\% \\
 High school diploma & 19 & 10.9\%  \\
 Two-year college degree / AA / AS &11 & 6.3\% \\
Four-year college degree / BA / BS & 69 & 39.4\% \\
Master's degree (e.g. MA, MS, MEd) & 28 & 16.0\% \\
Professional Degree (e.g. MD, DDS) & 25 & 14.3\% \\
Doctorate degree (e.g. PhD, EdD) & 22 & 12.6\%  \\

\hline
\end{tabular}
\caption{\textbf{Demographic information for the 175 study participants.}}\label{tab:demo}
 \vspace{-1em}
\end{table}

Each participant was remunerated with a reward of \$10 via an e-gift card. 10\% and 5\% of participants were randomly chosen to receive an additional \$25 and \$50, respectively.  We did not award additional money for correct answers so as not to incentize cheating (e.g., with Google reverse image search).  


\subsection{Knowledge Assessment Results}
We designed the knowledge assessment questions to range in difficulty, which was confirmed by the range of participant performances across questions.  The easiest question was answered correctly by 76\% of participants, and the hardest was answered correctly by 13.1\%.  The lower quartile of participant performance was 27.6\%, and the upper quartile was 60.4\%, indicating that the distribution was fairly balanced with a higher concentration of intermediate-level questions.

Of the 5 ``easy'' questions (defined as those with participant performance higher than the upper quartile), 2 were Logic, 2 were Math, and 1 was ML. Of the 12 ``intermediate'' questions (those with participant performance between the lower and upper quartiles), 1 was Logic, 6 were Math, and 6 were ML. Of the 5 ``hard'' questions (those with participant performance below the lower quartile), 1 was Math and 4 were ML. The following table maps each question to its difficulty level: 

\begin{table}[h]
\begin{tabular}{l|l} 
 \textbf{Easy} & L2, L3, M1, M2, ML1 \\
 \hline
 \textbf{Intermediate} & L1, M3, M4, M5, M6, M8, \\ 
 & ML2, ML3, ML4, ML5, ML6, ML10 \\ 
 \hline
 \textbf{Hard} & M7, ML7, ML8, ML9, ML11 \\
\end{tabular}
 \caption{Breakdown of questions across difficulty levels. The topics corresponding to each question are listed in Section 4.1.}
\end{table}

\begin{figure*}
\centering
   \begin{minipage}[!t]{0.47\textwidth}\centering%
   \captionsetup{width=\textwidth}
    \includegraphics[width=0.75\textwidth]{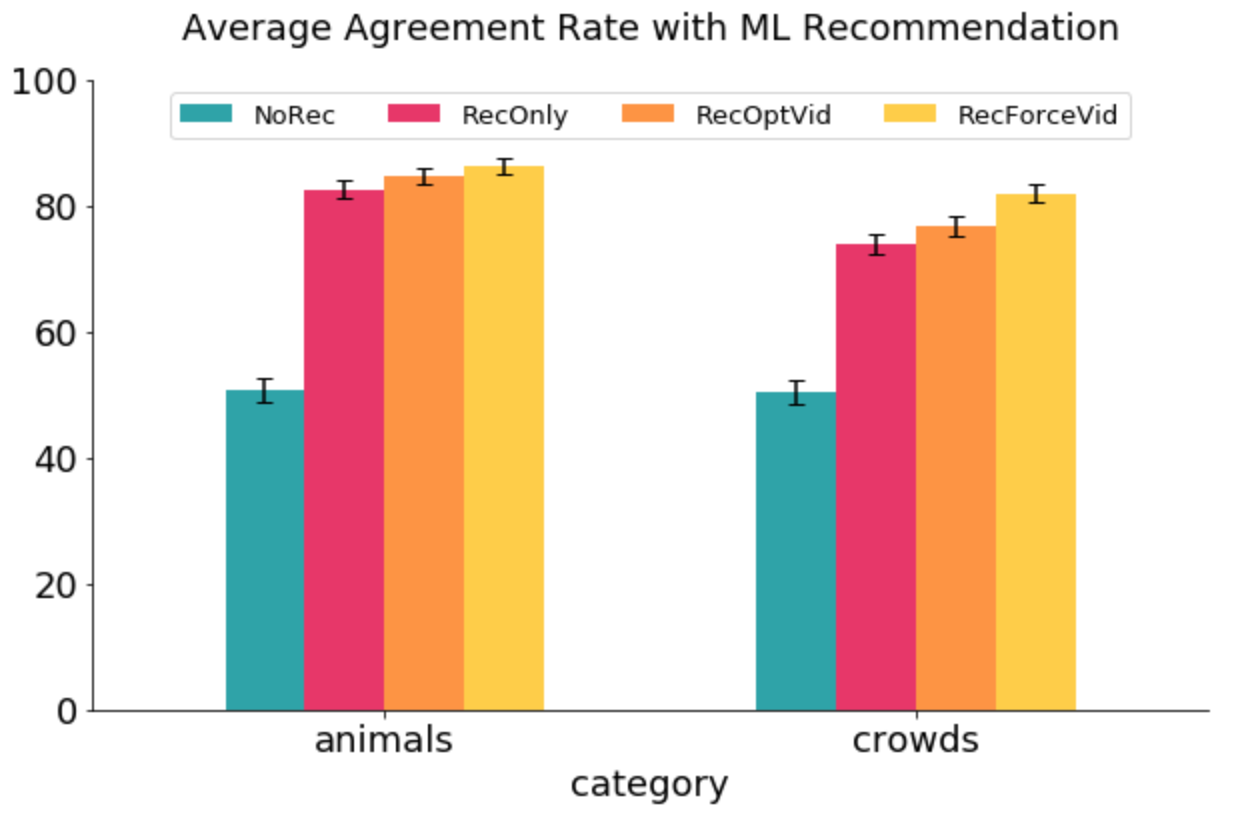}
  \caption{Percentage of questions where the ML recommendation was followed per treatment, for both applications. Error bars represent standard errors.}~\label{fig:Findings1}
   \end{minipage}%
   \hspace{\columnsep}
   \begin{minipage}[!t]{0.47\textwidth}\centering%
      \captionsetup{width=\textwidth}
     \includegraphics[width=0.8\textwidth]{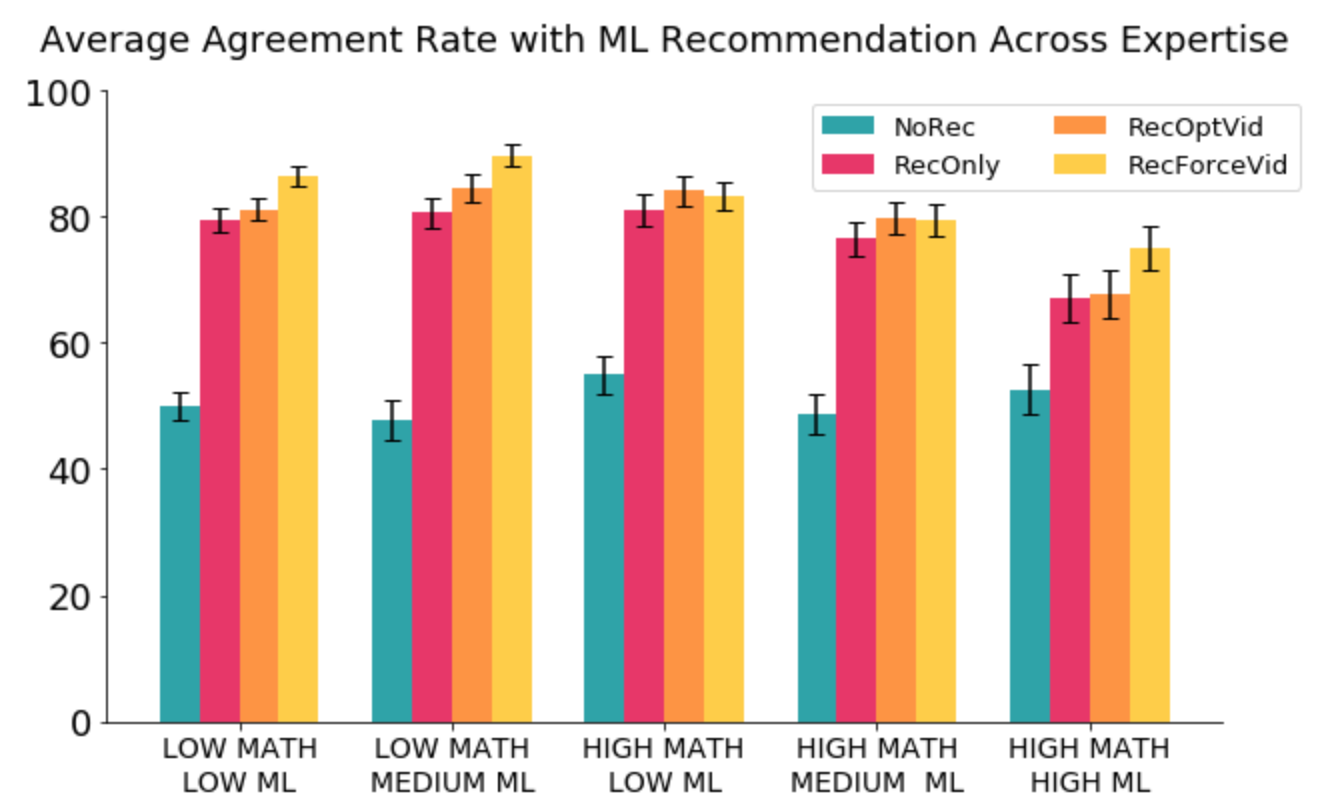}
      \caption{Percentage of questions where the ML recommendation was followed per treatment, per expertise group, for both applications combined. Error bars represent standard errors.}~\label{fig:FindingsA}
   \end{minipage}%
    \vspace{-1.5em}
\end{figure*}

\subsection{Expertise Assignment}
In the demographic survey, we asked participants to self-rank their ML background from 1-5 (1: No knowledge, 2: Novice, 3: Intermediate, 4: Advanced, 5: Expert) as a preliminary way to choose participants with varying knowledge. In analysis, we used participant responses to the knowledge assessment to more accurately assign expertise along two axes: Math/Logic (shorthand 'Math') and ML.  We gave each participant two scores corresponding to the percentage of the 11 Math/Logic asessment questions they answered correctly and the percentage of the 11 ML assessment questions they answered correctly. Scores were evenly distributed along the Math axis, so we set the boundary at 50\%, resulting in `Low Math' and `High Math'. ML scores were not as evenly distributed. We first set a `High ML' boundary to separate a small, distinct cluster of 19 participants with high ML scores ($>$60\%). We then set another boundary at the halfway point of the remaining scores (30\%), resulting in `Low ML' and `Medium ML'. The participants were thus categorized into five expertise groups: `Low Math-Low ML' (59), `Low Math-Medium ML' (33), `High Math-Low ML' (33), `High Math-Medium ML' (31), and `High Math-High ML' (19). 

We analyzed the differences between self-reported and assessment-assigned expertise groups and found that self-reported expertise was often unreliable, particularly for lower knowledge participants. People with Low ML assessment scores self-reported nearly evenly across all 5 ML expertise levels. In fact, there were more Low ML participants who self-reported as ``Experts'' than either the Medium or High ML groups. 
These discrepancies confirm that the implementation of a separate assessment was necessary to more accurately gauge people's ML expertise.

\section{Findings}
Participant responses to the task-based questions were analyzed using one-way within-subject ANOVA and the Tukey-Kramer test (All Pairs, Tukey HSD) across treatments. We used an alpha level of 0.05 for all statistical tests. All p-values are significant for the number of independent pair-wise comparisons using the Bonferroni correction.  Our findings are summarized as follows, and detailed in the coming subsections:
\begin{itemize}[topsep=0pt,itemsep=0ex,partopsep=1ex,parsep=1ex]
    \item \textbf{People generally follow ML recommendations}. This was the case across all expertise groups, and more so if the recommendations came with information about the ML system in the form of videos.
    \item \textbf{People follow incorrect ML recommendations for tasks they predominantly complete correctly, including experts}. 
    \item Although participants followed incorrect recommendations in general, they were followed significantly less often than correct recommendations, and \textbf{incorrect-abnormal recommendations were followed significantly less than incorrect-normal recommendations}.
    \item \textbf{Neither model, data, performance, nor prediction information improved people's accuracy}, even for incorrect-abnormal recommendations where the prediction video or image characteristics suggested that the recommendation should be less trustworthy.
\end{itemize}


\begin{figure*}
   \begin{minipage}[!t]{0.47\textwidth}\centering%
   \captionsetup{width=\textwidth}
    \includegraphics[width=0.8\textwidth]{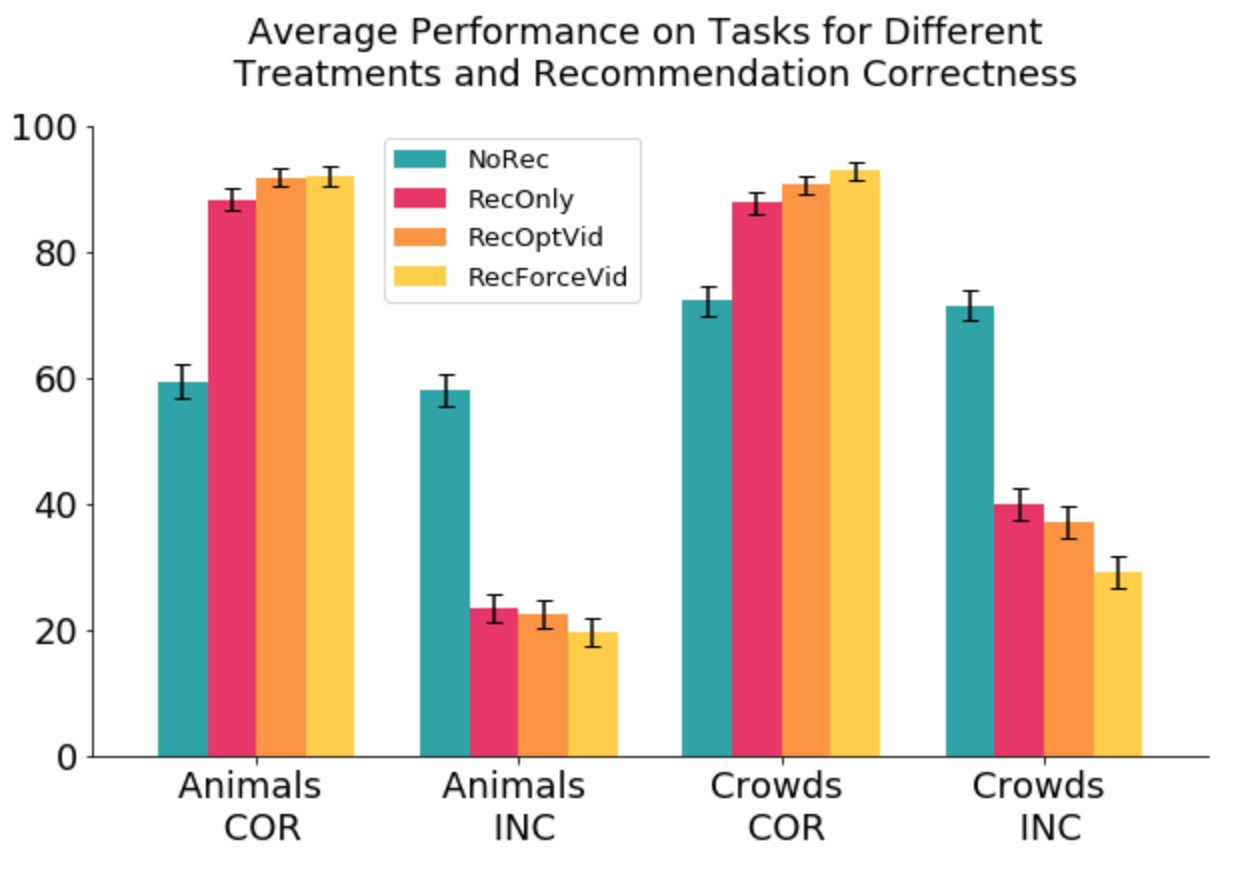}
  \caption{Performances per treatment, per application, separated by the correctness of the ML recommendation (COR = correct, INC = incorrect). Error bars represent standard errors.}~\label{fig:Findings2}
   \end{minipage}%
   \hspace{\columnsep}
   \begin{minipage}[!t]{0.47\textwidth}\centering%
      \captionsetup{width=\textwidth}
     \includegraphics[width=0.8\textwidth]{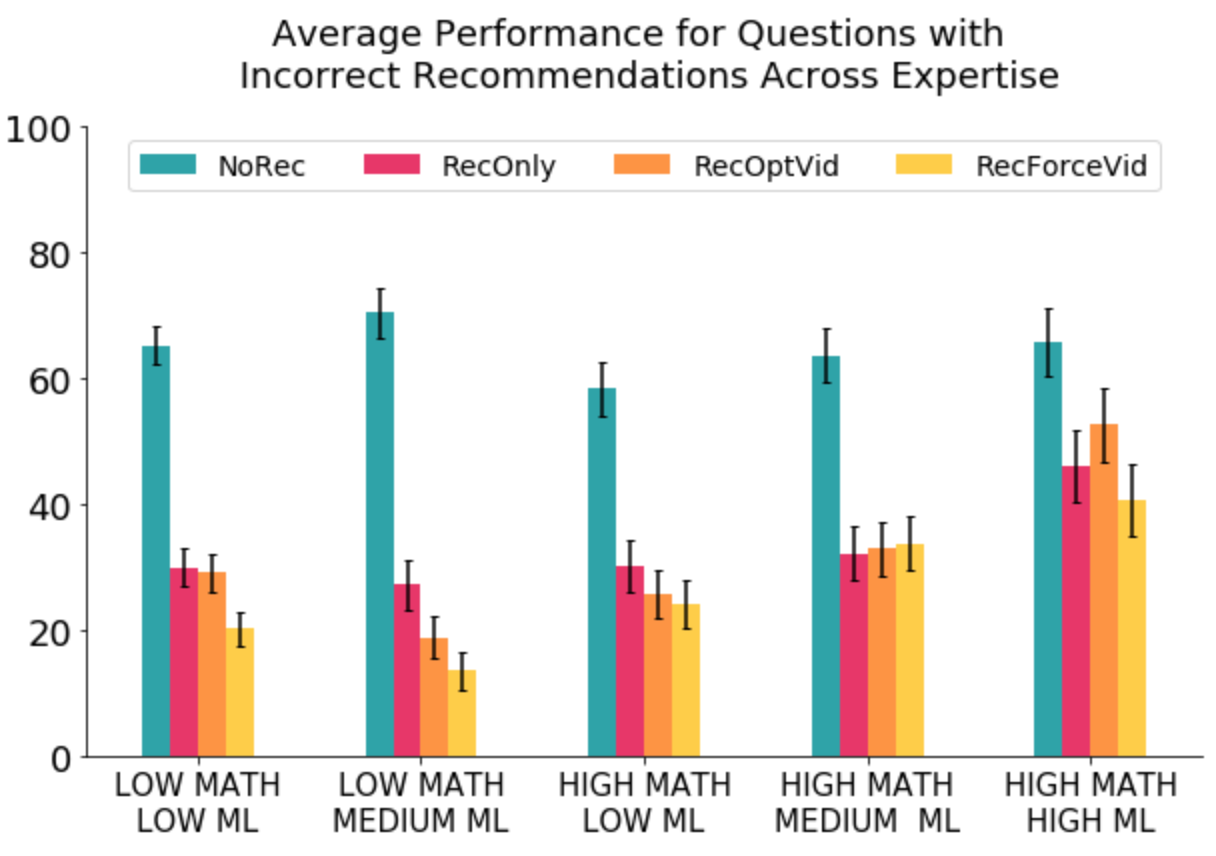}
      \caption{Performances for questions with incorrect recommendations per treatment, per expertise group, for both applications combined. Error bars represent standard errors.}~\label{fig:FindingsB}
   \end{minipage}%
    \vspace{-1.5em}
\end{figure*}

\subsection{People generally follow recommendations}
Overall analysis of participant responses revealed that people generally follow ML recommendations when they are provided, across treatments and applications. Additionally, participants who were given more information about the ML system and the recommendation in treatments RecOptVid and RecForceVid followed the recommendation more often than participants who only received the ML recommendation (treatment RecOnly) for both applications (Fig. \ref{fig:Findings1}).

\textit{Animal Identification:} Participants with the NoRec treatment answered the same as the hidden recommendations 50.7\% of the time. However, when participants were shown ML recommendations, they answered the same as the recommendation 84.4\% of the time on average. There were statistically significant differences in agreement between NoRec and each of the other treatments that showed ML recommendations [ANOVA: F(3, 696) = 107.48, \textit{p} $<$ 0.0001; Tukey HSD: \textit{p} $<$ 0.0001 for all (NoRec, \{RecOnly, RecOptVid, RecForceVid\}) pairs]. Recommendation agreement increased 2.2\% from RecOnly to RecOptVid and 3.7\% from RecOnly to RecForceVid. These increases were not statistically significant.

\textit{Crowd Comparison:} Participants with the NoRec treatment answered the same as the hidden recommendations 50.4\% of the time, while participants who were shown ML recommendations answered the same as the recommendations 77.5\% of the time. There were statistically significant differences in agreement between NoRec and each of the other treatments [F(3, 696) = 62.90, \textit{p} $<$ 0.0001; \textit{p} $<$ 0.0001 for all (NoRec, \{RecOnly, RecOptVid, RecForceVid\}) pairs]. We observe an increase of 2.8\% from RecOnly to RecOptVid and an increase of 8\% from RecOnly to RecForceVid. Only the RecOnly and RecForceVid treatment pair had a statistically significant difference [\textit{p} = 0.0073].

\textit{By Expertise:} There were statistically significant increases in agreement rate between NoRec questions and questions where recommendations were shown for every expertise group (Fig. \ref{fig:FindingsA}) [Low Math-Low ML: F(1, 116) = 142.54, \textit{p} $<$ 0.0001; Low Math-Medium ML: F(1, 64) = 100.45, \textit{p} $<$ 0.0001; High Math-Low ML: F(1, 64) = 48.91, \textit{p} $<$ 0.0001; High Math-Medium ML: F(1, 60) = 64.64, \textit{p} $<$ 0.0001; High Math-High ML: F(1, 36) = 12.48, \textit{p} = 0.0011].

\subsection{Incorrect recommendations are convincing}
Analysis of participant responses across questions with correct vs. incorrect ML recommendations revealed that people are willing to go against their intuition to follow a recommendation even if they complete the task correctly the majority of the time without a recommendation (Fig. \ref{fig:Findings2}); this trend is true even for ML experts. 

\textit{Animal Identification:} Participants answered questions correctly 58.7\% of the time overall without a recommendation. There was no statistically significant difference in performance across questions with correct versus incorrect recommendations for participants who did not receive recommendations (59.4\% vs. 58\%), indicating that these questions did not inherently vary in difficulty. However, for questions where participants were given the correct recommendation, participants performed on average 31.3\% better than in NoRec, which was statistically significant [F(3, 696) = 69.95, \textit{p} $<$ 0.0001; \textit{p} $<$ 0.0001 for all (NoRec, \{RecOnly, RecOptVid, RecForceVid\}) pairs]. For questions where participants were given the incorrect recommendation, they performed 36.1\% worse than in NoRec, which was also statistically significant [F(3, 696) = 52.22, \textit{p} $<$ 0.0001; \textit{p} $<$ 0.0001 for all (NoRec, \{RecOnly, RecOptVid, RecForceVid\}) pairs].

\textit{Crowd Comparison:} Participants answered questions correctly 71.9\% of the time overall without a recommendation. There was no statistically significant difference in performance for the NoRec treatments for questions with correct vs. incorrect recommendations (72.3\% vs. 71.4\%). However, for questions where participants were given the correct recommendation, they performed 18.1\% better than in NoRec, which was statistically significant [F(3, 696) = 24.46, \textit{p} $<$ 0.0001; \textit{p} $<$ 0.0001 for all (NoRec, \{RecOnly, RecOptVid, RecForceVid\}) pairs]. For questions where participants were given the incorrect recommendation, they performed 36\% worse than in NoRec, which was statistically significant [F(3, 696) = 47.12, \textit{p} $<$ 0.0001; \textit{p} $<$ 0.0001 for all (NoRec, \{RecOnly, RecOptVid, RecForceVid\}) pairs].

The baseline NoRec performance for the crowd comparison application (71.9\%) was significantly higher than the baseline for the animal identification application (58.7\%), but the final accuracies for questions where the correct recommendation was given are approximately the same and extremely high for both tasks (90.7\% for animal identification and 90.4\% for crowd comparison). The decreases in performance after receiving incorrect recommendations were also approximately equal for both tasks (36.1\% for animal identification and 36\% for crowd comparison).

\textit{By Expertise:} There were no statistically significant differences between the baseline NoRec performances across the expertise groups. However, there were statistically significant decreases in performance from NoRec to each of the RecOnly/RecOptVid/RecForceVid treatments for all expertise groups except for High Math-High ML, which only had statistical significance from NoRec to RecForceVid (Fig. \ref{fig:FindingsB}). While not statistically signficant, receiving more information tends to decrease performance for the lower expertise groups, while for the higher expertise groups this trend is not present (High Math-High ML) or is slightly reversed (High Math-Medium ML).


Although participants in all expertise groups still followed incorrect recommendations overall, participants in the High Math-High ML group performed statistically significantly better than participants in all the other groups except for High Math-Medium ML for questions where incorrect recommendations were shown [F(4, 170) = 4.48, \textit{p} = 0.0018; \textit{p} = 0.0109 for (*, Low Math-Low ML), \textit{p} = 0.0009 for (*, Low Math-Medium ML), \textit{p} = 0.0276 for (*, High Math-Low ML)]. 

Prior math knowledge also seemed to have more effect than ML knowledge on performance for questions with incorrect recommendations that showed videos. For participants with Medium ML knowledge, those who had High Math knowledge performed statistically significantly better than their Low Math counterparts for the RecForceVid treatment [F(4, 170) = 5.34, \textit{p} = 0.0005; \textit{p} = 0.0096 for (High Math-Medium ML, Low Math-Medium ML)], even though there was no significant difference in these groups for the RecOnly treatment. There were no statistically significant differences in performance when comparing across Low ML and Medium ML, holding Math knowledge constant.

\subsection{Abnormal recommendations are followed less}
\begin{figure}
    \centering
    \includegraphics[width=0.8\columnwidth]{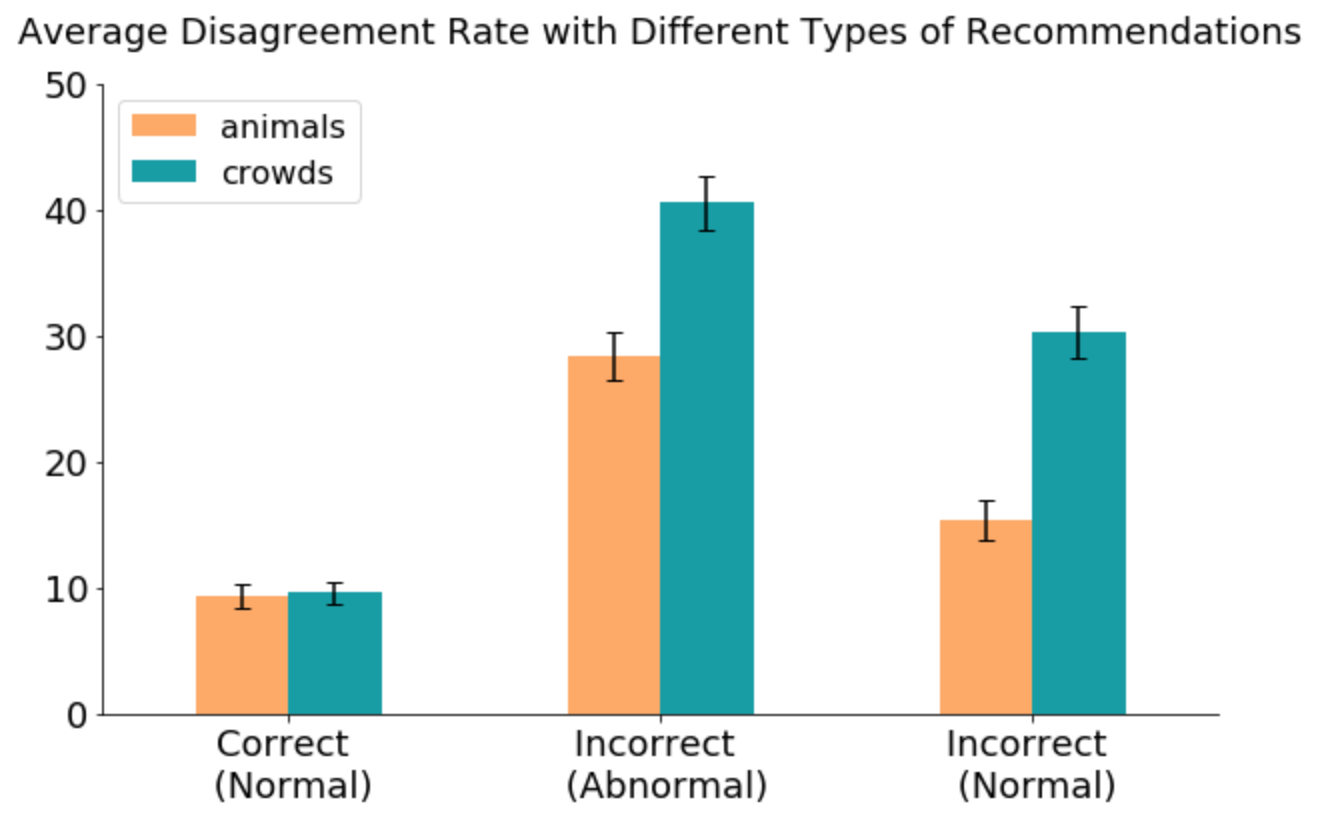}
    \caption{Percentage of questions where participants disagreed with a shown recommendation per application, per the 3 recommendation types. Error bars represent standard errors.}~\label{fig:INCABNORM}
     \vspace{-1.6em}
\end{figure}

Participants disagreed with shown correct recommendations 9.5\% of the time, and with shown incorrect recommendations 28.7\% of the time. The incorrect recommendations designated as ``abnormal'' also had a significantly higher rate of recommendation disagreement than the incorrect recommendations designated as ``normal'' for both applications (Fig. \ref{fig:INCABNORM}). 

\textit{Animal Identification:} Participants did not follow a correct recommendation 9.3\% of the time, whereas they did not follow an incorrect recommendation 21.9\% of the time --- split between 28.4\% for incorrect-abnormal recommendations and 15.4\% for incorrect-normal recommendations. There were statistically significant differences between all three of these values [F(2, 522) = 29.14, \textit{p} $<$ 0.0001; \textit{p} $<$ 0.0001 for (COR\_NORM, \{INC\_ABNORM, INC\_NORM\}) pairs, \textit{p} = 0.0450 for (INC\_ABNORM, INC\_NORM)]. 

\textit{Crowd Comparison:} Participants did not follow a correct recommendation 9.6\% of the time, whereas they did not follow an incorrect recommendation 35.4\% of the time on average --- split between 40.6\% for incorrect-abnormal recommendations and 30.3\% for incorrect-normal recommendations. There were statistically significant differences between all three of these values [F(2, 522) = 50.51, \textit{p} $<$ 0.0001; \textit{p} $<$ 0.0001 for (COR\_NORM, \{INC\_ABNORM, INC\_NORM\}) pairs, \textit{p} = 0.0032 for (INC\_ABNORM, INC\_NORM)]. 

It is interesting to note that while participants disagreed with recommendations at similar rates when given correct recommendations in the animal identification and crowd comparison tasks (9.3\% and 9.6\%), the crowd comparison task had significantly higher disagreement rates than the animal identification task for both incorrect-abnormal recommendations [F(1, 348) = 10.83, \textit{p} = 0.0011] and incorrect-normal recommendations [F(1, 348) = 25.20, \textit{p} $<$ 0.0001]. This difference between the two applications may be correlated to the baseline difficulty of the task (crowd comparison: 71.9\% NoRec accuracy, animal identification: 58.7\% NoRec accuracy).

\subsection{Information does not improve accuracy}

In this study, we used videos about the ML system and recommendation as a way to provide transparency. As we saw in Figs. \ref{fig:Findings2} and \ref{fig:FindingsB}, offering videos simply led to participants following the ML recommendation more, and did not improve participant performance for questions with incorrect recommendations. For incorrect recommendations, watching explanatory videos did not improve performance over the NoRec baselines for either task or any expertise groups. For participants who watched the performance videos, there was also no correlation between the stated performance of the model on a subtask, and participants' trust of the model for that subtask. Performances ranged from 72\% to 87\% for animal identification, and from 72\% to 93\% for crowd comparison. 

We also analyzed the rate at which participants followed recommendations based on whether or not they had watched each of the 4 types of videos. Across all 3 recommendation types and all 4 video types, watching any video increased the recommendation agreement rate, even for incorrect-abnormal recommendations where the video about prediction information showed extremely low ML probabilities (Fig. \ref{fig:vids}, bottommost). 

\textit{Correct Recommendations:} There were statistically significant increases in recommendation agreement between the not-watched and watched-in-question states for the performance videos [F(2, 500) = 4.87, \textit{p} = 0.0080; \textit{p} = 0.0071] and prediction videos [F(1, 348) = 7.60, \textit{p} = 0.0061]. There were observable increases in agreement for the data and model videos as well from not-watched to watched-in-question. 

\textit{Incorrect-Normal Recommendations:} There were statistically significant increases in recommendation agreement between the not-watched and watched-in-question states for all videos. The data videos had F(2, 370) = 3.85, \textit{p} = 0.0222; \textit{p} = 0.0264]. The model videos had F(2, 378) = 3.87, \textit{p} = 0.0218; \textit{p} = 0.0224]. The performance videos had F(2, 449) = 6.44, \textit{p} = 0.0017; \textit{p} = 0.0012. The prediction videos had F(1, 348) = 12.44, \textit{p} = 0.0005. 

\textit{Incorrect-Abnormal Recommendations:} There were no statistically significant differences between watch states for any of the videos, but we observed this increasing pattern between not-watched and watched-in-question for all 4 videos. 

\begin{figure}
\centering
\subfloat[] {\includegraphics[width=.3\textwidth]{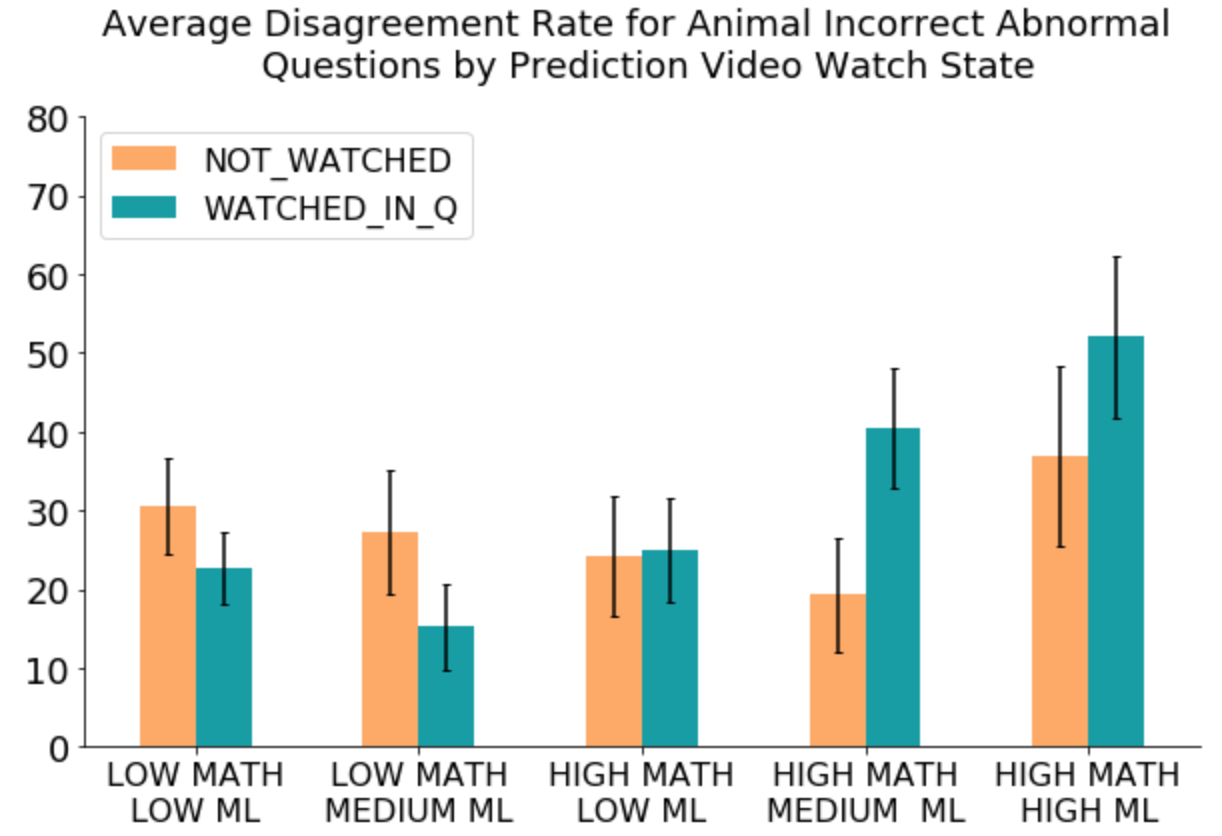}}

\subfloat[] {\includegraphics[width=.3\textwidth]{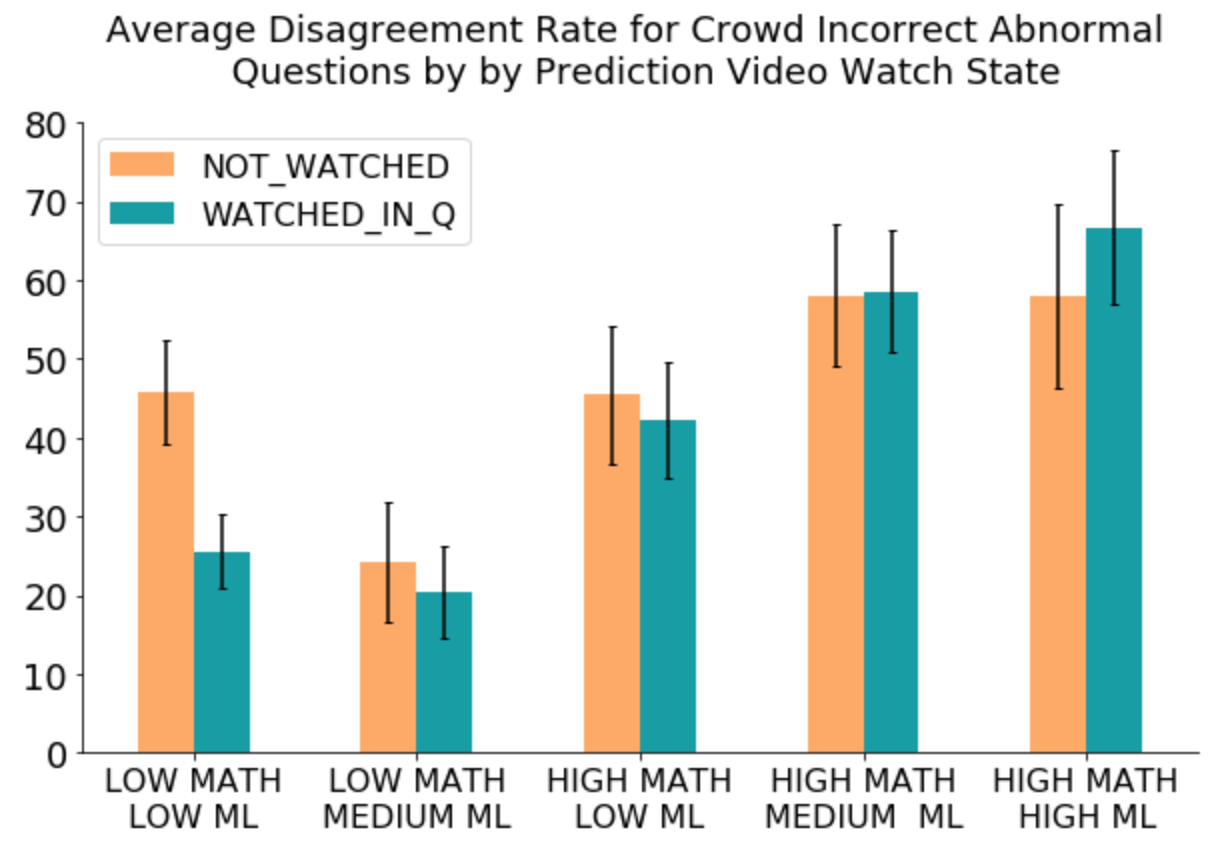}}
~
\caption{Percentage of questions where participants disagreed with a shown incorrect-abnormal recommendation per task where (a) is animal identification and (b) is crowd comparison, per watch state of the prediction video, per expertise group. Error bars represent standard errors.} \label{fig:predvids}
 \vspace{-1em}
\end{figure}

\textit{By Expertise:} Interestingly, participants in higher Math and ML knowledge groups who watched prediction videos disagreed more with the incorrect-abnormal recommendations than if they had not watched the videos. Lower knowledge groups agreed more after watching those same videos (Fig. \ref{fig:predvids}).

\textit{Animal Identification:} For questions that showed incorrect-abnormal recommendations, there were decreases in recommendation disagreement rate from not-watched to watched for both Low Math groups. However, there were \textit{increases} in recommendation disagreement rate from not-watched to watched for the High Math-Low ML, High Math-Medium ML group and the High Math-High ML group. The only statistically significant difference was in the High Math-Medium ML group [F(1, 60) = 4.28, \textit{p} = 0.0428].

\textit{Crowd Comparison:} For questions that showed incorrect-abnormal recommendations, there were decreases in recommendation disagreement rate from not-watched to watched for both Low Math groups and for the High Math-Low ML group. However, there were \textit{increases} in recommendation disagreement rate from not-watched to watched for the High Math-Medium ML group and the High Math-High ML group. The only statistically significant difference was in the Low Math-Low ML group [F(1, 116) = 8.03, \textit{p} = 0.0054].

The positive differences between not-watched and watched were higher in the animal identification application than in the crowd comparison application. This may be because incorrect-abnormal examples in the animal application had very low, close probabilities (Fig. \ref{fig:vids}, bottom), while the abnormalities in the crowd comparison application were more subtle, e.g., a poorly lit image. 

\section{Limitations}
The remote and web-based nature of the study is limited; e.g., we were not able to ensure that participants were paying attention throughout the study. Although the participants could not switch screens while videos were playing, we could not be sure that they were focused on the screen.  Additionally, the tasks we present are scenarios described online, not in-person scenarios with real consequences. As a result, trust and behavior in real scenarios might differ from what we observe in this study.  

\section{Discussion}
Our study contributes three key insights into human-ML interaction and trust: 

1. \textbf{People trust ML recommendations when they should not}. In our study, we found that even when people (a) are able to perform the task correctly the majority of the time without an ML recommendation, (b) are given information that points to very low system confidence, and (c) have the expertise to understand it, they still follow an incorrect ML recommendation. This result agrees with prior work suggesting that people over-rely on ML recommendations \cite{logg2019algorithm,povyakalo2013discriminate,skitka2000accountability}, and builds upon it by confirming this claim across five different math/logic/ML expertise levels. Ideally, ML recommendations should augment human performance by providing support for decisions where people would otherwise make mistakes; in conjunction, humans should be able to recognize signals indicating that a recommendation is less trustworthy, and then rely more on their own judgement. Our findings show some promise in that people disregard incorrect recommendations more often than correct ones. However, incorrect recommendations, even those that suggest low trustworthiness, are still overwhelmingly followed. This finding strongly motivates continued research into effective ways to communicate a model's limitations.

2. \textbf{Relevant, curated system information should be prioritized over quantity to avoid over-reliance}. Other work has studied the effect of system information on trust in specific domains or with narrow categories of information \cite{yin2019understanding,bussone2015role,Sinha:2002:RTR:506443.506619,Cramer2008}, and we build upon this with similar findings for four different types of information. Policy-makers are pushing for increased transparency into ML systems, and people affected by these systems deserve to have access to this information as well. However, simply providing comprehensive system/prediction information is not sufficient, and may wrongly lead people to follow incorrect recommendations. The type of information that will be helpful to real users will be different for each application and needs to be developed through deeply engaging with end-users. Moreover, figuring out the best way to communicate this information to users is an equally necessary area of study.

3. \textbf{Math and logic skills may be as important as ML skills for people who need to interpret the results of ML systems}. Many current resources for learning ML are aimed at ML practitioners, computer scientists, or those with an existing technical background. Although few of these resources are understandable to the general public, ML will certainly be used by people from a range of educational backgrounds and occupations. Currently, many educators who teach ML to non-technical audiences actively avoid teaching with math altogether \cite{teachmeml}. However, our findings motivate the importance of developing resources that teach the necessary math and logic skills in concurrence with ML skills specifically for these non-technical audiences.

This work is a step towards ML systems that work together with human users to deliver the positive impact they promise.

\begin{acks}
Harini Suresh and Ilaria Liccardi were funded by the MIT-IBM Watson AI Lab. Ilaria Liccardi was also supported by the William and Flora Hewlett Foundation.
\end{acks}

\balance{}
\bibliographystyle{ACM-Reference-Format}
\bibliography{sample}

\end{document}